\documentclass[aps,pre,reprint,groupedaddress]{revtex4-1}
\usepackage{amsmath}
\usepackage{graphicx}
\usepackage{bm}
\usepackage{dcolumn}
\usepackage[colorlinks,linkcolor=blue,anchorcolor=black,citecolor=blue,urlcolor=blue,bookmarks=false]{hyperref}
\usepackage{epstopdf}

\begin{document}


\title{Do the Navier-Stokes equations embody all physics in a flow of Newtonian fluids?}


\author{Qifeng Lv}
\email[]{keephong@126.com}
\author{Sijing Wang}
\affiliation{Department of Hydraulic Engineering, Tsinghua University, Beijing 100084, China}



\begin{abstract}
The Navier-Stokes equations are the governing equations of fluid flows. They are deemed to embody all physics in a flow of Newtonian fluids like water, especially when we assume the fluid is incompressible. Fluid flows are usually described in the Eulerian perspective wherein we focus on the fluid's quantities at a fixed point in space as time processes, rather than the motion of each individual particle. The latter perspective is indeed the Lagrangian perspective which is widely used in solid mechanics like elasticity. Thus, the Navier-Stokes equations should be and indeed they are always used in the Eulerian perspective. Here we show the right-hand sides of the Navier-Stokes equations were derived not in the Eulerian perspective but rather in the Lagrangian perspective, because the strain rates used in the derivation were the Lagrangian Cauchy strain rates. Thus, for describing fluid flows in the Eulerian perspective, the Navier-Stokes equations may lost some physics. To make it sure, we put the Cauchy strain into the material derivative, and then derive the Eulerian Cauchy strain rates. We find the difference between the Eulerian and the Lagrangian Cauchy strain rates is significant when in a turbulent flow. Therefore, On the basis of the Eulerian Cauchy strain rates, we derive a new set of governing equations for the flow of Newtonian fluids. The newly derived governing equations are fully in the Eulerian perspective. We believe they are the real equations that Navier, Cauchy, Poisson and Stokes intended to find for the viscous flow of Newtonian fluids.
\end{abstract}


\maketitle

\textbf{Introduction}. -- As the governing equations of fluid flows, the Navier-Stokes equations are still full of mysteries. The main reason for the mysteries perhaps is the mathematical difficulty in solving these equations \cite{L'vov1998}. In fact, to find a singular solution of the Navier-Stokes equations in the three dimensional space \cite{Leray1934} is one of the seven Millennium Prize Problems proposed by the Clay Mathematics Institute \cite{Fefferman2000,*Dickson2000}. The difficulty is all because of the nonlinearity, and the nonlinearity also leads these equations to be unstable. Thus, we may reasonably believe these unstable equations control the chaotic behaviors of turbulence which is also of great interest and mysteries for centuries \cite{Hof2006,*Eckhardt2009,*Avila2011}. Moreover, the Navier-Stokes equations' connection with Einstein's general relativity also fascinates many scientists \cite{Bredberg2012,*Green2014}. Although the mysteries make the Navier-Stokes equations popular in many fields, they may also veil some drawbacks of the Navier-Stokes equations.

\begin{figure*}
  \includegraphics[width=16cm]{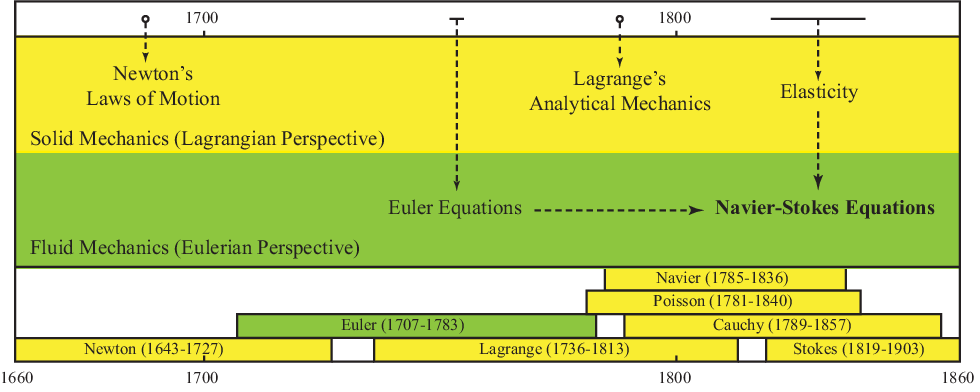}\\
  \caption{(Color online) The two foundations of the Navier-Stokes equations: Euler equations and elasticity. \textbf{Note that the two foundations are in different perspectives.}}\label{fig1}
\end{figure*}

Now, let us unveil the mysterious equations step by step. The nonlinear terms in the Navier-Stokes equations seem to contribute all the mysteries. However, the nonlinear terms arise not from the physical attributes of the fluids but rather from kinematics. In fact, the nonlinear terms in the Navier-Stokes equations are the same as those in the Euler equations which, derived by Euler \cite{Euler1755a} in 1755, are nonlinear kinematical equations. The Euler equations are used to describe the inviscid fluid flows, and they are one of the two foundations for the Navier-Stokes equations. (The other foundation is elasticity of solids. See fig.\ref{fig1}) We rewrite the Euler equations below by using the indicial notation:
\begin{equation}
\rho \left(\frac{\partial v_i}{\partial t}+v_j\frac{\partial v_i}{\partial x_j}\right)=f_{i}-\frac{\partial p}{\partial x_i},\label{eq1}
\end{equation}
where $\rho$ is the fluid's density, $v$ is the velocity, $t$ is the time, $x$ is the spatial coordinate, $f$ is the body force, $p$ is the pressure, and the subscripts $i$ and $j$, ranging in values from 1 to 3, represent the three components of a physical quantity in space. Whenever an indicial subscript occurs twice in the same term, we understand the subscript is to be summed from 1 to 3. For instance, $x_{ii}=\sum_{i=1}^{3}x_{ii}$. Thus, the index can be chosen freely. The nonlinear terms in the Euler equations, i.e., $v_j(\partial v_i/\partial x_j)$, are indeed quasi linear rather than fully nonlinear. These terms are due to the Eulerian perspective \cite{Euler1755a} wherein we focus on the fluid's quantities at a fixed point in space as time processes, rather than the motion of each individual particle. The latter perspective is the Lagrangian perspective \cite{Newton1687,*Lagrange1788} which is widely used in solid mechanics like elasticity.

Seventy years later after the derivation of the Euler equation, Navier \cite{Navier1822}, then Cauchy \cite{Cauchy1828} and Poisson \cite{Poisson1829}, considering the Euler equations are unable to describe the viscous flow, respectively derived another set of equations for the viscous flow from 1822 to 1829. These equations are just the Navier-Stokes equations which we call them today. In fact, The Navier-Stokes equations were named after Stokes \cite{Stokes1845} gave a clearly concluding derivation in 1845 that was almost a century later since the derivation of the Euler equations \cite{Euler1755a} (see fig.\ref{fig1}). The Navier-Stokes equations for incompressible fluids can be written as
\begin{equation}
\rho \left(\frac{\partial v_i}{\partial t}+v_j\frac{\partial v_i}{\partial x_j}\right)=f_{i}-\frac{\partial p}{\partial x_i}+\mu\frac{\partial^2v_i}{\partial x_j^2},\label{eq2}
\end{equation}
where $\mu$ is the fluid's viscosity which is a constant for Newtonian fluids. The difference between the Euler equations and the Navier-Stokes equations is the viscous terms which contain the viscosity $\mu$ in equation (\ref{eq2}). This difference is attributed to the viscous stresses which Navier, Cauchy, Poisson and Stokes derived respectively by analogy with elasticity of solids.

\begin{figure*}
  \centering
  \includegraphics[width=16cm]{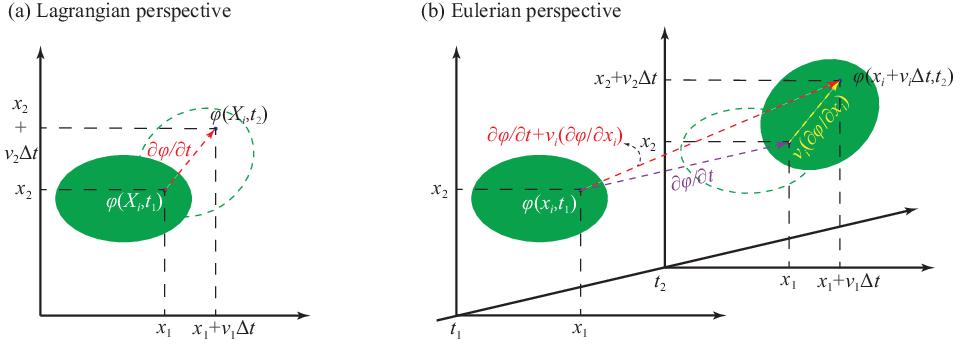}\\
  \caption{(Color online) The difference between the time rates of a physical quantity in the (a) Lagrangian perspective and the (b) Eulerian perspective. $\phi$ is a physical quantity, $t$ is the time, $x$ is the spatial coordinate, and the subscripts, $i$ and $j$, 1 and 2, represent the components of a physical quantity in space. In the Lagrangian perspective, the time rate of $\phi$ can be expressed as $\dot{\phi}=\partial \phi/\partial t$, cf. eq. (\ref{eq3}) and eq. (\ref{eq4}), whereas the time rate in the Eulerian perspective is $\dot{\phi}={\rm D}\phi/{\rm D}t=\partial \phi/\partial t+v_i(\partial \phi/\partial x_i)$, cf. eq. (\ref{eq5}) and eq. (\ref{eq6}).}\label{fig2}
\end{figure*}

Without doubt, the Euler equations were derived fully in the Eulerian perspective. On the basis of the Euler equations, the Navier-Stokes equations are also always used in the Eulerian perspective. In the following section of this Letter, we show the Navier-Stokes equations were not fully derived in the Eulerian perspective because the strain rates used in the derivation were indeed derived in the Lagrangian perspective. These strain rates are actually today's Lagrangian Cauchy strain rates. We thus derive a new form of the Cauchy strain rates in the Eulerian perspective and then find the difference between the two Cauchy strain rates cannot be neglected in turbulent flows or compressible fluid flows. Therefore, using the present Eulerian Cauchy strain rates and other assumptions in \cite{Stokes1845}, we finally derive the governing equations for the momentum conservation of Newtonian fluids in this Letter.

\textbf{Cauchy strain rates in the Eulerian perspective}. -- Before starting to derive the Cauchy strain rates in the Eulerian perspective, let us see the difference for a physical quantity $\phi$'s time rates $\dot{\phi}$  between the two perspectives (i.e., the Lagrangian perspective and the Eulerian perspective). See fig. \ref{fig2}. By Taylor's theorem and neglecting higher order terms when $\Delta t \to 0$, we have the expression below for $\dot{\phi}$ in the Lagrangian perspective
\begin{eqnarray}
\dot{\phi}\Delta t&=&\phi\left(X_i,t+\Delta t\right)-\phi\left(X_i,t\right)\nonumber
\\
 &=&\phi+\frac{\partial\phi}{\partial t}\Delta t-\phi,\label{eq3}
\end{eqnarray}
where $X$ represents the coordinate of an individual particle. Thus, from eq. (\ref{eq3}), the time rate of $\phi$ in the Lagrangian perspective can be derived, namely
\begin{equation}
\dot{\phi}=\frac{\partial\phi}{\partial t}.\label{eq4}
\end{equation}
This is indeed the most familiar time derivative in solid mechanics.

For fluid mechanics, we often use the Eulerian perspective. In the Eulerian perspective, similarly but differently, we have
\begin{eqnarray}
\dot{\phi}\Delta t&=&\phi\left(x_i+v_i\Delta t,t+\Delta t\right)-\phi\left(x_i,t\right)\nonumber
\\
 &=&\phi+\frac{\partial\phi}{\partial t}\Delta t+\frac{\partial\phi}{\partial x_i}v_i\Delta t-\phi.\label{eq5}
\end{eqnarray}
Then $\dot{\phi}$ in the Eulerian perspective can be expressed as
\begin{equation}
\dot{\phi}=\frac{\partial\phi}{\partial t}+v_i\frac{\partial\phi}{\partial x_i}=\frac{{\rm D}\phi}{{\rm D}t},\label{eq6}
\end{equation}
where the operator ${\rm D}/{\rm D} t=\partial/\partial t+v_i(\partial/\partial x_i)$ is the material derivative. Comparing eq. (\ref{eq4}) with eq. (\ref{eq6}), we are clear that the expressions for the time rate of a physical quantity in the two perspectives differ from each other.

Coming back to our question, from \cite{Stokes1845}, we know that the strains used in the Navier-Stokes equations are the Cauchy strains although they were not having such a name during the time of Navier and Stokes. The Cauchy strains are evolved from the analysis of elasticity, which can be written as
\begin{equation}
\epsilon_{ij}=\frac{1}{2}\left(\frac{\partial u_i}{\partial x_j}+\frac{\partial u_j}{\partial x_i}\right),\label{eq7}
\end{equation}
where $u$ is the displacement. Then the strain rates used in the derivation of the Navier-Stokes equations are expressed as
\begin{equation}
\dot{\epsilon}^{\rm L}_{ij}=\frac{1}{2}\left(\frac{\partial v_i}{\partial x_j}+\frac{\partial v_j}{\partial x_i}\right).\label{eq8}
\end{equation}
Note that the displacement $u$ in eq. (\ref{eq7}) is changed into the velocity $v$ in eq. (\ref{eq8}). This is correct in the Lagrangian perspective because $\dot{u}_i=\partial u_i/\partial t=v_i$, and this is indeed the most familiar derivation in Newtonian or Lagrangian mechanics \cite{Lagrange1788,*Newton1687} which are widely used to describe solids' behaviors. However, for fluids in the Eulerian perspective, this may lost something because
\begin{equation}
\dot{u}_i=\frac{{\rm D}u_i}{{\rm D}t}=\frac{\partial u_i}{\partial t}+v_j\frac{\partial u_i}{\partial x_j}=v_i.\label{eq9}
\end{equation}
Given this, we may reasonably conjecture that Navier, Cauchy, Poisson and Stokes did not notice the difference between the Lagrangian perspective and the Eulerian perspective. After all, Euler had passed away half a century before the derivation of the Navier-Stokes equations (see fig.\ref{fig1}). On the other hand, we should also be clear that some physical quantities having the same time-rate expression in the two perspectives may exist. However, nobody has ever claimed the Cauchy strains are such quantities.

To check it out, we substitute the Cauchy strains, i.e. eq. (\ref{eq7}), into the material derivative. That is
\begin{eqnarray}
\dot{\epsilon}_{ij}&=&\frac{{\rm D}\epsilon_{ij}}{{\rm D}t}\nonumber
\\
  &=&\frac{1}{2}\left(\frac{\partial v_i}{\partial x_j}+\frac{\partial v_j}{\partial x_i}-\frac{\partial v_k}{\partial x_j}\frac{\partial u_i}{\partial x_k}-\frac{\partial v_k}{\partial x_i}\frac{\partial u_j}{\partial x_k}\right).\label{eq10}
\end{eqnarray}
Now we can find the Eulerian Cauchy strain rates $\dot{\epsilon}_{ij}$ differ from $\dot{\epsilon}^{\rm L}_{ij}$ which, used in the Navier-Stokes equations, are the Lagrangian Cauchy strain rates. The difference is the last two terms in eq. (\ref{eq10}). When in high-Reynolds-number flows or compressible-fluid flows, both $\partial v_i/\partial x_j$ and $\partial u_i/\partial x_j$ are large. Therefore, we cannot neglect the last two terms of eq. (\ref{eq10}) in these cases.

\textbf{Governing equations for the flow of Newtonian fluids}. -- Both the Euler equations and the Navier-Stokes equations are governing equations for the momentum conservation of fluid flows.
 The left-hand sides of these two equations are just the mass-times-acceleration of Newton's second law expressed in the Eulerian perspective, and the right-hand sides represent forces, which should also be in the Eulerian perspective. The forces are actually the stresses which correlate to the strain rates by certain principles. In the above, we have derived the Cauchy strain rates fully in the Eulerian perspective, which are different from those used in the Navier-Stokes equations. Therefore, the governing equations for the momentum conservation of fluid flows, which depend on the strain rates, should also be in a different appearance.

To derive the governing equations fully in the Eulerian perspective, following \cite{Stokes1845}, we only consider Newtonian fluids here. That is to say the shear stress $\tau$ correlates to the strain rate by $\tau=\mu\dot{\epsilon}$, and the stress tensor is expressed as $\sigma_{ij}=-p\delta_{ij}+\lambda\dot{\epsilon}_{kk}\delta_{ij}+2\mu\dot{\epsilon}_{ij}$, where $\lambda$ is the second viscosity and $\delta_{ij}$ is the Kronecker delta ($\delta_{ij}=1$, when $i=j$; $\delta_{ij}=0$, when $i\neq j$). Substituting $\dot{\epsilon}_{ij}$ into $\sigma_{ij}$, we have
\begin{eqnarray}
\sigma_{ij}&=&-p\delta_{ij}+\lambda\left(\frac{\partial v_m}{\partial x_m}-\frac{\partial v_n}{\partial x_m}\frac{\partial u_m}{\partial x_n}\right)\delta_{ij}\nonumber
\\
 & &+\mu\left(\frac{\partial v_i}{\partial x_j}+\frac{\partial v_j}{\partial x_i}\right.\nonumber
\\
 & &\left.-\frac{\partial v_k}{\partial x_j}\frac{\partial u_i}{\partial x_k}-\frac{\partial v_k}{\partial x_i}\frac{\partial u_j}{\partial x_k}\right).\label{eq11}
\end{eqnarray}
The momentum conservation of a fluid (i.e., Newton's second law expressed in the Eulerian perspective) can be expressed as
\begin{equation}
\rho \left(\frac{\partial v_i}{\partial t}+v_j\frac{\partial v_i}{\partial x_j}\right)=f_{i}+\frac{\partial\sigma_{ij}}{\partial x_j}.\label{eq12}
\end{equation}
Substituting $\dot{\epsilon}_{ij}$ and $\sigma_{ij}$ into eq. (\ref{eq12}), we obtain the governing equations for the flow of Newtonian fluids fully in the Eulerian perspective, namely
\begin{eqnarray}
\rho \left(\frac{\partial v_i}{\partial t}+v_j\frac{\partial v_i}{\partial x_j}\right)&=&f_{i}-\frac{\partial p}{\partial x_i}+\left(\lambda+\mu\right)\left(\frac{\partial^2 v_j}{\partial x_i\partial x_j}\right.\nonumber
\\
 & &\left.-\frac{\partial^2 v_k}{\partial x_i\partial x_j}\frac{\partial u_j}{\partial x_k}-\frac{\partial v_k}{\partial x_j}\frac{\partial^2 u_j}{\partial x_i\partial x_k}\right)\nonumber
\\
 & &+\mu\left(\frac{\partial^2 v_i}{\partial x_j^2}-\frac{\partial^2 v_k}{\partial x_j^2}\frac{\partial u_i}{\partial x_k}\right.\nonumber
\\
 & &\left.-\frac{\partial v_k}{\partial x_j}\frac{\partial^2 u_i}{\partial x_i\partial x_j}\right).\label{eq13}
\end{eqnarray}
If we assume the fluid is incompressible, i.e.,
\begin{subequations}
\label{eq14}
\begin{eqnarray}
\epsilon_{ii}&=&\frac{\partial v_i}{\partial x_i}=0\label{eq14a}
\\
\dot{\epsilon}_{ii}&=&\frac{\partial v_i}{\partial x_i}-\frac{\partial v_i}{\partial x_j}\frac{\partial u_j}{\partial x_i}=0,\label{eq14b}
\end{eqnarray}
\end{subequations}
then, eq. (\ref{eq13}) can be reduced to
\begin{eqnarray}
\rho \left(\frac{\partial v_i}{\partial t}+v_j\frac{\partial v_i}{\partial x_j}\right)&=&f_{i}-\frac{\partial p}{\partial x_i}+\mu\left(\frac{\partial^2 v_i}{\partial x_j^2}-\frac{\partial^2 v_k}{\partial x_j^2}\frac{\partial u_i}{\partial x_k}\right.\nonumber
\\
 & &\left.-\frac{\partial v_k}{\partial x_j}\frac{\partial^2 u_i}{\partial x_i\partial x_j}\right).\label{eq15}
\end{eqnarray}
Equation (\ref{eq15}) is just the governing equations for the flow of incompressible Newtonian fluids, which can be closed by eqs. (\ref{eq9}) and (\ref{eq14}). Comparing eq. (\ref{eq15}) with the Navier-Stokes equations, i.e., eq. (\ref{eq2}), we find the difference is the last two terms in eq. (\ref{eq15}). According to the above analysis for the Cauchy strain rates, we know these two terms cannot be neglected in turbulent flows.

\textbf{Discussion}. -- We will discuss and explain some mysteries and controversies by using the newly derived governing equations for the flow of Newtonian fluids.

For the Millennium Prize Problem on the singular solutions of the Navier-Stokes equations, we can, anyway, assume the solutions are regular and do not blow up. If this is true, many scientists then conjectured there must be an unknown magical mechanism to help the viscosity control the nonlinearity \cite{Kerr2011}. On the other hand, if the Navier-Stokes solutions do blow up, that will also bring problems because we have never experienced an infinite velocity in nature or experiments. In this case, singularity (blowup) is deemed missing essential physics \cite{Kerr2011}. Therefore, whether the solutions blow up or not, the Navier-Stokes equations will possibly breakdown in some cases. Moreover, Foias et al. showed great curiosity on why the nonlinearity of the governing equations for viscous flows is not contributed by physical attributes \cite{Foias2001}. In fact, the newly derived governing equations for the incompressible flow of Newtonian fluids, i.e., eq. (\ref{eq15}), may shed some light on these mysteries. In eq. (\ref{eq15}), we indeed find the mechanism lost by the Navier-Stokes equations. This mechanism is nonlinear, does arise from the fluids' physical attributes and cannot be neglected when in a turbulent flow.

For the hydrodynamic stability, Lin pointed out that the existing theory to explain the instability of high-Reynolds-number flows is confused and thus revising the Navier-Stokes equations is a possible necessity \cite{Lin1944}. Moreover, he also found the viscosity indeed plays a rather complicated role in a flow. It not only damps out the disturbance but also cause the instability \cite{Lin1955}. Indeed, this fact can be well explained by the present governing equations. At the right-hand side of eq. (\ref{eq15}), the terms containing $\mu$ just show the dual role of viscosity. Moreover, eq. (\ref{eq15}) is in fact an answer to Lin's conjecture and proposal for revising the Navier-Stokes equations.

For compressible fluids, ever since 2004 Brenner has believed the Navier-Stokes equations lost something essential \cite{Brenner2004}, and he thus proposed another hydrodynamics to replace the Navier-Stokes-Fourier paradigm \cite{BrennerInt2012}. The present equation (\ref{eq12}), which are the governing equations for compressible Newtonian fluids, may provide some support to Brenner's revision because the newly appeared terms in equation (\ref{eq12}), which are indeed rigorously derived, cannot be neglected either for compressible fluids. Therefore, a revision of the Navier-Stokes equations is needed in this situation.

\textbf{Conclusion}. -- The Eulerian perspective is the foundation to describe fluid flows. However, we find that the right-hand sides of the Navier-Stokes equations were derived from the Lagrangian perspective because the strain rates used in the derivation are the Lagrangian Cauchy strain rates. Given the fact that many mysteries and controversies arose on the Navier-Stokes equations when faced turbulent flows or compressible fluid flows, we predict the deviation arising from neglecting the difference of the two perspectives is significant in such cases although it may be trivial in the incompressible laminar flow. In fact, after deriving the Cauchy strain rates in the Eulerian perspective, we find the difference between the Lagrangian and the Eulerian Cauchy strain rates is distinct. Thus, on the basis of the Eulerian Cauchy strain rates, we derive the governing equations for the flow of Newtonian fluids fully in the Eulerian perspective. The newly derived governing equations are fully nonlinear partial differential equations. This attribute actually unraveled some mysteries arising before on the Navier-Stokes equations. Moreover, it is reasonable to believe the present governing equations will bring more fresh insights into the mysterious turbulence.

\textbf{Acknowledgments}. -- Commemorate Mr. Howard Brenner. Thanks for the discussions, Stokes' paper (1845) and all the help.

\end{document}